\documentclass[preprint,amsmath,amssymb,nobibnotes,aps,prx,floatfix]{revtex4-2}

\usepackage{graphicx}
\usepackage{dcolumn}
\usepackage{bm}

\def\Comp{\mathcal{\hat{C}}}
\def\diff{{\mathrm{d}}}
\newcommand{\RRC}[1]{\mathcal{R}(#1)}
\DeclareMathOperator\arctanh{arctanh}
\def\Cah{\hat{\mathrm{Ca}}}

\begin{document}

\preprint{APS/123-QED}

\title{Dynamics of compression-driven gas--liquid displacement in a capillary tube}

\author{Callum Cuttle}
\author{Christopher W. MacMinn}%
\email{christopher.macminn@eng.ox.ac.uk}
\affiliation{Department of Engineering Science, University of Oxford, Oxford, OX1 3PJ, UK}

\date{\today}

\begin{abstract}
We study two-phase displacement via the steady compression of an air reservoir connected to an oil-filled capillary tube. Our experiments and modelling reveal complex displacement dynamics depending on compression rate and reservoir volume that, for large reservoirs, depend on a single dimensionless compressibility number. We identify two distinct displacement regimes, separated by a critical value of the compressibility number. While the subcritical regime exhibits quasi-steady displacement after an initial transient, the supercritical regime exhibits burst-like expulsion.
\end{abstract}

\maketitle

The gas-driven displacement of viscous liquid from a confined geometry occurs in various natural and industrial systems, including subsurface storage of carbon dioxide~\citep{Berg2012,Zhao2016}, operation of fuel cells~\citep{Lee2019}, reopening of airways~\citep{Hazel2003,Ducloue2017}, and displacement of subretinal blood during eye surgery~\citep{Pappas2021}. Gas--liquid displacement is one of the simplest classes of two-phase flows because the viscosity of the gas is typically negligible, making these flows especially tractable to theoretical analysis and firmly establishing their role as the key model system in interfacial fluid dynamics~\citep[\textit{e.g.},][]{Casademunt2004}. Nonetheless, gas-driven displacements are inherently unsteady due to the interaction between compression of the gas and viscous resistance in the liquid. For example, recent work has identified non-monotonic variations in pressure or invasion rates during displacement in porous media~\citep{Louriou2011,Lee2019}, episodic growth of intricate patterns during displacement of granular suspensions~\citep{Sandnes2011,Sandnes2012}, and time-dependent growth of fractures during injection of foams into gels~\citep{Lai2018}, despite the constant nominal injection rate in all cases. Here, we show that even the simplest gas-driven displacement is a complex dynamical system where spring-like compression drives flow against a rate- and state-dependent resistance.

\begin{figure*}
  \includegraphics[width=1\linewidth]{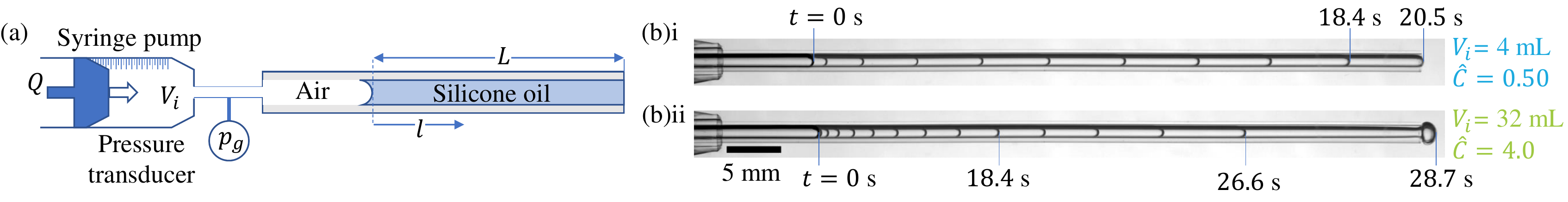}
  \caption{ \label{fig:set_up} Displacement of silicone oil from a capillary tube by the injection of air from a reservoir of initial volume $V_i$ that is compressed at a constant volume rate $Q$ using a syringe pump. (a)~Experimental set-up. We measure the displacement of the interface $l(t)$ relative to its initial position [$l(0)=0$] and the gauge pressure $p_g$ of the air. The oil slug has initial length $L$. (b)~Experimental timelapses of overlaid frames at equal timesteps $\Delta t=2.05\pm0.02$~s show the motion of the interface; interframe spacing of interfaces is inversely proportional to interface velocity. (i) and (ii) are for $V_i=4$~mL and 32~mL, respectively, and $Q=0.2$~mL/min ($\Comp=0.50$ and 4.0; see also Fig.~\ref{fig:res1}).}
  \end{figure*}

To elucidate the nonlinear dynamics underpinning these flows, we consider a model problem: displacement of viscous oil from a capillary tube by compressing a connected reservoir of air of initial volume $V_i$ at a fixed nominal rate $Q$ [Fig.~\ref{fig:set_up}(a)]. The fluid mechanics of this idealised system are fully captured by a simple model that reproduces our experimental observations both qualitatively and quantitatively, as discussed below. Yet, complex behavior emerges immediately from both experiments and theory: At the same $Q$, an experiment can either tend to a steady velocity after an initial transient when $V_i$ is small [Fig.~\ref{fig:set_up}(b)i] or accelerate rapidly toward a burst-like expulsion when $V_i$ is large [Fig.~\ref{fig:set_up}(b)ii]. Here, we show that these examples illustrate the two distinct displacement regimes that emerge from the coupling of spring-like compression with viscous displacement. We show that this problem is a specific realisation of a simple dynamical system, the complex behaviour of which can be captured with a reduced model comprising a nonlinear, first-order ordinary differential equation. This model reproduces and explains the contrasting dynamics shown in Fig.~\ref{fig:set_up}b, as well as the sharp transition point between these two dynamical regimes.

Our experimental flow cell comprised a glass capillary tube of length 10~cm and inner radius $R=0.66\pm0.01$~mm [Fig.~\ref{fig:set_up}(a)]. One end of the tube was connected to a sealed air reservoir of initial volume $V_i\in\{4,8,16,32\}\pm0.1$~mL; the other end fed into a bath of silicone oil at fixed hydrostatic pressure $p_{HS}$ (viscosity $\mu=0.096$~Pa~s, surface tension $\gamma=21$~mN~m$^{-1}$, and density $\rho=960$~kg~m$^{-3}$ at laboratory temperature $T_{\mathrm{lab}}=22\pm1~^{\circ}$C; Dow Corning). Before each experiment, oil was drawn from the bath into the tube, filling an initial length $L=56\pm1$~mm. Experiments were initiated by compressing the air reservoir at a fixed rate $Q\in\{0.05,0.1,0.2,0.4,0.8,1.6\}$~mL/min, thus injecting air into the tube and expelling oil into the bath. We used imaging and image processing to measure the motion of the interface, the thickness of thin residual films deposited on the tube walls by the perfectly wetting oil, and the radius of curvature $b$ of the air-oil interface. We also measured the gauge pressure $p_g$ of the air relative to atmospheric pressure $p_\mathrm{atm}=101$~kPa, and thus calculated the viscous pressure drop $\Delta p=p_g-2\gamma/b-p_{HS}$ across the oil slug (see Appendix~\citep{see_supp}).

\begin{figure*}
\includegraphics[width=0.9\linewidth]{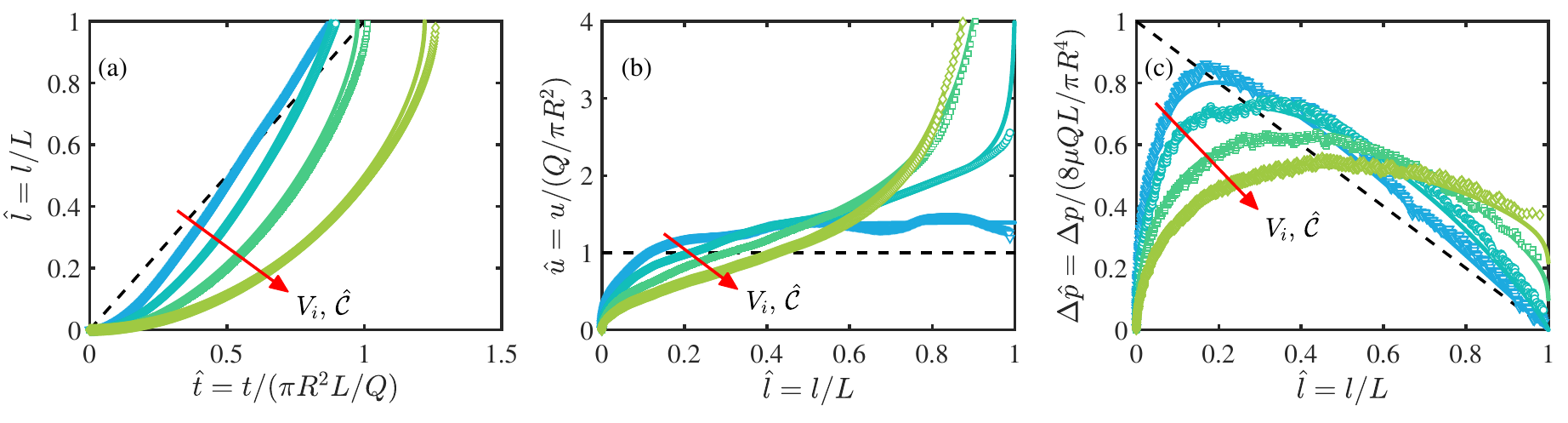}
\caption{ \label{fig:res1} (a-c)~Experimental results (symbols) and numerical solutions to the full model with films (solid lines; see Appendix~\citep{see_supp}) for $Q=0.2$~mL/min and $V_i=4,8,16$~and~32~mL ($\Comp=0.50$, 1.0, 2.0 and 4.0; dark to light). (a)~Normalised displacement $\hat{l}$ of the interface as a function of normalised time $\hat{t}$. (b-c)~Normalised velocity $\hat{u}$ and normalised pressure drop $\Delta\hat{p}$ as functions of $\hat{l}$. Dashed lines show corresponding incompressible behaviour for reference. Arrows indicate increasing $V_i$ and $\Comp$.}
\end{figure*}

For reference, we first consider an incompressible displacement at a rate $Q$ (dashed lines in Fig.~\ref{fig:res1}). Neglecting thin residual films, the air-oil interface must advance linearly with time $t$, such that its displacement relative to its initial position is $l = Qt/(\pi R^2)$ [Fig.~\ref{fig:res1}(a)] and its velocity $u=\diff l/\diff t =Q/(\pi R^2)$ is constant [Fig.~\ref{fig:res1}(b)]. The experiment ends at breakout time $t_{bo}=\pi R^2L/Q$, when the interface reaches the outlet (\textit{i.e.}, $l(t_{bo})=L$). Assuming laminar Hagen-Poiseuille flow (see below), the pressure gradient in the oil $\Delta p/(L-l)=8\mu Q/(\pi R^4)$ is constant and the pressure drop $\Delta p$ decreases linearly from its maximum value of $8\mu QL/(\pi R^4)$ (reached instantaneously after flow starts at $t=0$) to its breakout value of $\Delta p_{bo}=0$ [Fig.~\ref{fig:res1}(c)]. We normalise our compressible results below by these incompressible reference values, such that $\hat{l}=l/L$, $\hat{t}=t/(\pi R^2L/Q)$, $\Delta\hat{p}=\Delta p/(8\mu QL/(\pi R^4))$ and $\hat{u}=u/(Q/(\pi R^2))$. Hatted quantities throughout are dimensionless.

\begin{figure}
\includegraphics[width=0.7\linewidth]{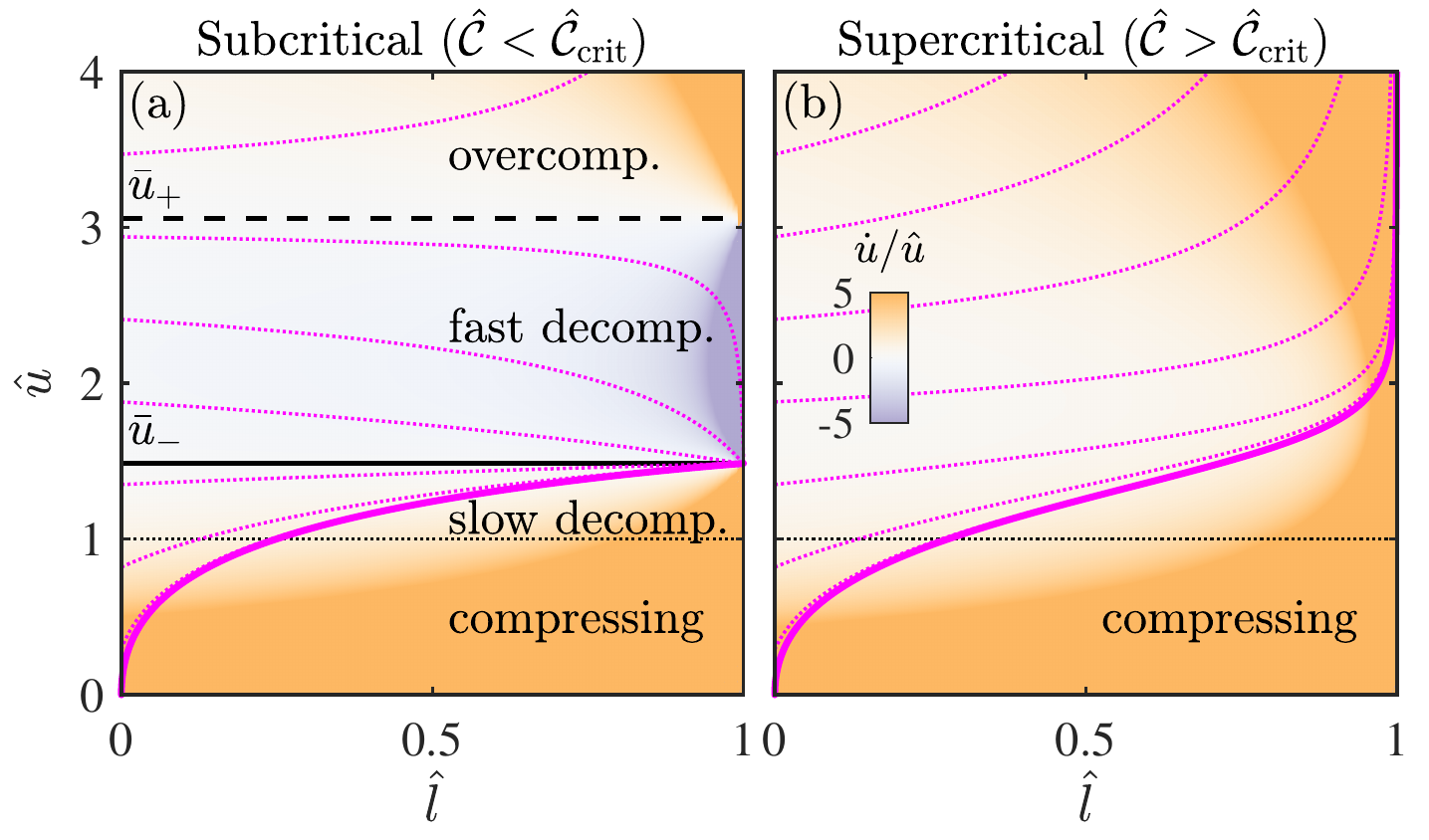}
\caption{ \label{fig:res3} Phase diagram in $\hat{u}$-$\hat{l}$ space for (a) sub- and (b) supercritical expulsion dynamics ($\Comp=0.88$ and 1.12, respectively). Example trajectories with varying $\hat{u}(0)$ are plotted as magenta lines; the thick trajectory has $\hat{u}(0)=0$, as in our experiments. Dashed, solid and dotted black lines indicate $\bar{u}_\pm$ and $\hat{u}=1$, respectively. The colormap indicates $\dot{u}/\hat{u}$ (see colorbar), which is saturated in regions close to $\hat{l}=1$ and $\hat{u}=0$.}
\end{figure}

In our experiments, the interface advances nonlinearly in time [Fig.~\ref{fig:res1}(a)], initially moving slowly [$\hat{u}<1$; Fig~\ref{fig:res1}(b)] as the air compresses and pressurises [$\Delta\hat{p}$ increases; Fig.~\ref{fig:res1}(c)]. As oil drains and air pressure builds, $\hat{u}$ increases. Once $\hat{u}>1$, the oil flux exceeds the nominal flux imposed by the pump and the air begins decompressing [$\Delta\hat{p}$ decreases; Fig.~\ref{fig:res1}(c)]. (Ref.~\citep{Khemili2021} made similar observations, but across a more limited range of parameters.) As a result, the maximum in $\Delta\hat{p}$ occurs when $\Delta\hat{p}$ crosses the incompressible solution, for which $\Delta\hat{p}=1-\hat{l}$ and $\hat{u}=1$. As $V_i$ is increased (arrows in Fig.~\ref{fig:res1}), we observe the transition from quasi-steady to burst-like expulsion [\textit{e.g.}, Figs.~\ref{fig:set_up}(b)i and \ref{fig:set_up}(b)ii, respectively]. This transition is most apparent in the normalised velocities measured close to breakout [$\hat{u}(\hat{l}\approx1)$], which increase dramatically from slightly above to more than 10 times the nominal velocity as $V_i$ increases [not visible on the scale of Fig.~\ref{fig:res1}(b); see Fig.~\ref{fig:res3}(c)]. We next introduce a simple model that captures these observations.

We model the air as a fixed mass of isothermal ideal gas. We take the oil pressure at the outlet to be atmospheric, so the initial absolute air pressure is $p_\mathrm{atm} + 2\gamma/R$. We neglect $p_{HS}\ll p_{atm}$ as it is arbitrary, but retain $2\gamma/R\ll p_{atm}$ for generality as it depends on the system parameters. The gauge pressure of the air is then
\begin{equation}\label{eq:relPres}
p_g(t)=\left(p_\mathrm{atm}+2\gamma/R\right)\frac{V_i}{V(t)}-p_\mathrm{atm},
\end{equation}
where $V(t)$ is the current volume of air and $V(0)=V_i$. The syringe pump acts to decrease $V$ at a steady rate $Q$, while the motion of the interface acts to increase $V$ at a rate $\pi R^2 (\diff l/\diff t)$; hence,
\begin{equation}\label{eq:AirVol}
V(t)=V_i-Qt+\pi R^2l(t).
\end{equation}
We model the oil flow as Hagen-Poiseuille flow. In the absence of thin films, the interface velocity must be equal to the mean oil velocity, such that
\begin{equation}\label{eq:HPvel}
\frac{\diff l}{\diff t}=\frac{R^2}{8\mu}\left(\frac{\Delta p}{L-l}\right),
\end{equation}
where the viscous pressure drop along the oil slug is $\Delta p=p_g-2\gamma/R$. Substituting Eqs.~(\ref{eq:relPres})~and~(\ref{eq:AirVol}) into Eq.~(\ref{eq:HPvel}) and introducing $\hat{l}$, $\hat{t}$, and $\Delta\hat{p}$ from above yields
\begin{equation}\label{eq:ODEnondim}
\frac{\diff \hat{l}}{\diff\hat{t}}= \left[\frac{\hat{p}_0+2/\hat{\mathrm{Ca}}}{\hat{V_i}+\left(\hat{l}-\hat{t}\right)}\right]\left(\frac{\hat{t}-\hat{l}}{1-\hat{l}}\right)=\frac{\Delta\hat{p}}{\left(1-\hat{l}\right)}.
\end{equation}
Equation~(\ref{eq:ODEnondim}) is a nonlinear ordinary differential equation containing three independent nondimensional parameters: $\hat{p}_0=\pi R^4p_\mathrm{atm}/(8\mu QL)$, comparing the compressive and viscous pressure scales; the capillary number $\hat{\mathrm{Ca}}=8\mu QL/(\pi R^3 \gamma)$, comparing the viscous and capillary pressure scales; and $\hat{V_i}=V_i/(\pi R^2L)$, comparing the initial volumes of air and oil.

To quantitatively compare with experiments requires the inclusion of thin films, which modify the interfacial capillary pressure and the kinematic relation between interface and oil velocities. We do so in the Appendix~\citep{see_supp} using well-established results and corrections. The full model with thin films (solid lines in Fig.~\ref{fig:res1}) is in strong quantitative agreement with the experiments with no fitting parameters. As demonstrated below, however, the key qualitative features of this system can be captured with a much simpler model that does not include thin films.

Our model can be simplified by considering the limit of a much larger initial volume of air than of oil, $\hat{V_i}\gg1$, which is the case in our experiments ($\hat{V_i}\approx50-420$). In this limit, which corresponds to approximating the air as a linear spring, Eq.~(\ref{eq:ODEnondim}) reduces to
\begin{equation}\label{eq:ODEred}
\frac{\diff \hat{l}}{\diff \hat{t}}\approx\left[\frac{\hat{p}_0+2/\hat{\mathrm{Ca}}}{\hat{V_i}}\right]\left(\frac{\hat{t}-\hat{l}}{1-\hat{l}}\right)\equiv \frac{4}{\Comp}\left(\frac{\hat{t}-\hat{l}}{1-\hat{l}}\right).
\end{equation}
The system is then governed by a single non-dimensional `compressibility number',
\begin{equation}\label{eq:Comp}
\Comp=\frac{32\mu QV_i}{\pi^2R^6p_\mathrm{atm}}\left(1+\frac{2\gamma}{R p_\mathrm{atm}}\right)^{-1}.
\end{equation}
The reduced model, Eq.~(\ref{eq:ODEred}), captures most features of the full model and the experiments, and permits an implicit analytical solution given in the Appendix~\citep{see_supp}. The compressibility number $\Comp$ can be interpreted by considering the characteristic rates at which compressive and viscous pressures vary, \textit{i.e.}, $\dot{P}_C=Qp_{atm}/V_i$ and $\dot{P}_V=8\mu Q^2/(\pi^2R^6)$, respectively. Comparing with Eq.~(\ref{eq:Comp}), we find that $\Comp\approx4(\dot{P}_V/\dot{P}_C)$ when $2\gamma/(R p_\mathrm{atm})\ll1$ (\textit{i.e.}, when the capillary pressure is much less than $p_\mathrm{atm}$), as is the case for $R\gtrsim10$~$\mu$m. Thus, $\Comp$ measures the rate of viscous depressurisation relative to compressive pressurisation.

\begin{figure}
\includegraphics[width=0.55\linewidth]{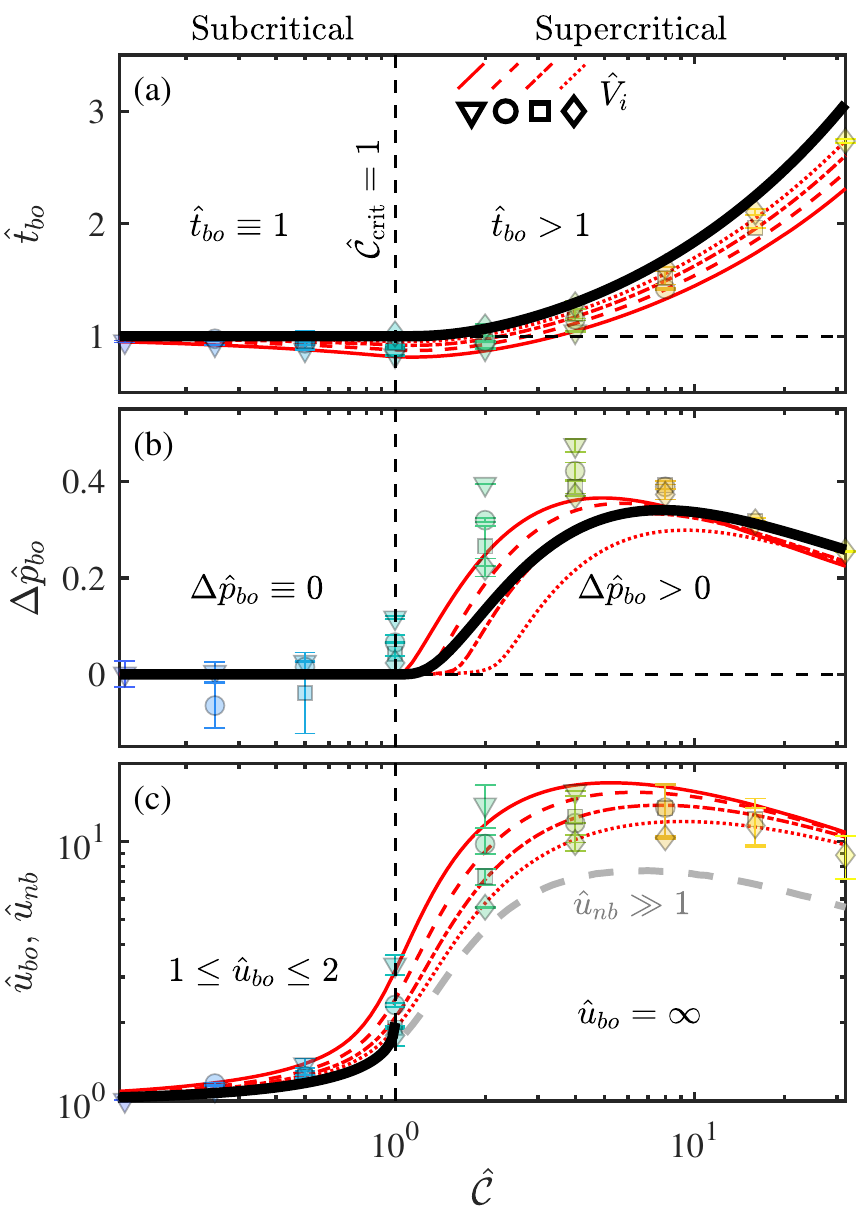}
\caption{ \label{fig:res2} Normalised breakout quantities calculated analytically from the reduced model (Eq.~\ref{eq:ODEred}). Thick black curves show (a) $\hat{t}_{bo}$, (b) $\Delta{\hat{p}_{bo}}$ and (c) $\hat{u}_{bo}$ as functions of $\Comp$. Vertical dashed lines at $\Comp=\Comp_\mathrm{crit}$ separate the sub- and supercritical regimes. In (c), the near-breakout velocity $\hat{u}_{nb}$ is plotted as a thick dashed curve. For comparison, $\hat{t}_{bo}$, $\Delta\hat{p}_{bo}$ and $\hat{u}_{nb}$ are also shown for experiments (symbols) and the full model with thin films (red curves; see Appendix~\citep{see_supp}) for different $\hat{V}_i$ [see legend in (a)].}
\end{figure}

The effect of varying $\Comp$ in the reduced model (Eq.~\ref{eq:ODEred}) is summarised in Fig.~\ref{fig:res2}, which shows (normalised) breakout time $\hat{t}_{bo}$ [Fig.~\ref{fig:res2}(a)], breakout pressure drop $\Delta\hat{p}_{bo}$ [Fig.~\ref{fig:res2}(b)], and breakout velocity $\hat{u}_{bo}$ [Fig.~\ref{fig:res2}(c)] as functions of $\Comp$. These breakout quantities all exhibit two distinct regimes separated by a critical compressibility number $\Comp_\mathrm{crit}=1$ (dashed lines). For all $\Comp\le\Comp_\mathrm{crit}$ (\textit{i.e.}, subcritical expulsion), breakout occurs at exactly $\hat{t}_{bo}\equiv1$, meaning that the time taken to drain the oil is identical to that of an incompressible displacement at the same $Q$. As a result, the volume of air displaced by the piston at the moment of breakout is exactly equal to the volume of oil expelled, so the air returns precisely to its initial volume and pressure (\textit{i.e.}, $\Delta\hat{p}_{bo}\equiv0$). For all $\Comp>\Comp_\mathrm{crit}$ (\textit{i.e.}, supercritical expulsion), breakout is delayed [$\hat{t}_{bo}>1$; Fig.~\ref{fig:res3}(a)]. The volume displaced by the piston at the moment of breakout is greater than the volume of oil expelled, so the air is compressed and the system terminates with an overpressure, $\Delta\hat{p}_{bo}>0$ [Fig.~\ref{fig:res3}(b)]. These two scenarios have dramatically different consequences for the breakout velocity, $\hat{u}_{bo}=\lim_{\hat{l}\to1}\Delta\hat{p}/(1-\hat{l})$, which can only remain finite if $\Delta\hat{p}$ tends to zero at the same rate that $\hat{l}$ tends to 1. This is the case only for subcritical expulsion, in which $1\le\hat{u}_{bo}\le2$. During supercritical expulsion, the overpressure at breakout results in infinite breakout velocities. Note that our use of the terms subcritical and supercritical refers only to the value of $\hat{\mathcal{C}}$ relative to $\hat{\mathcal{C}}_\mathrm{crit}$, and is not intended to imply anything about the nature of the bifurcation. See Table~S1 of Appendix~\citep{see_supp} for analytical expressions for breakout quantities in each regime.

Measurements of breakout time $\hat{t}_{bo}$, breakout pressure $\Delta\hat{p}_{bo}$ and near-breakout velocity $\hat{u}_{nb}$ are also plotted in Fig.~\ref{fig:res2}(a-c) for experiments and for the full model with films for different values of $\hat{V}_i$. We measure $\hat{u}_{nb}$ two tube diameters from the outlet [$\hat{u}_{nb}=\hat{u}(\hat{l}=1-4R/L)$] to reduce uncertainty by avoiding large $\hat{u}_{bo}$ at high $\Comp$. Our experimental results clearly reflect a transition between sub- and supercritical regimes at $\Comp=\Comp_\mathrm{crit}$, consistent with the predictions of the reduced model (Eq.~\ref{eq:ODEred}). For $\Comp<\Comp_\mathrm{crit}$, $\Delta\hat{p}_{bo}\approx0$ to within experimental uncertainty [Fig.~\ref{fig:res2}(b)], suggesting the pressure driving the flow vanishes at breakout, and the breakout velocities are modest [$\hat{u}_{nb}\gtrsim1$; Fig.~\ref{fig:res2}(c)]. For $\Comp>\Comp_\mathrm{crit}$, we observe a marked rise in both $\Delta\hat{p}_{bo}$ and $\hat{u}_{nb}$, consistent with an overpressure driving burst-like expulsions with $\hat{u}_{nb}\sim10$. The spread in the data and in the predictions of the full model for different $\hat{V}_i\propto V_i$ at fixed $\Comp\propto Q V_i$ is due to residual films, the thickness of which depends strongly on $\hat{\mathrm{Ca}}\propto Q$.

The reduced model (Eq.~\ref{eq:ODEred}) describes a dynamical system in which the steady compression of a linear spring drives motion against a state-dependent viscous damper. To rationalise the emergence of such complex behavior from this simple, first-order dynamical system, we consider the generalised system in $\hat{u}$-$\hat{l}$ phase space, as depicted in Fig.~\ref{fig:res3}. Introducing the resistance $\hat{\omega}=1-\hat{l}$, Eq.~(\ref{eq:ODEred}) can be written as
\begin{equation}\label{eq:rates}
\RRC{\hat{u}}=\RRC{\Delta\hat{p}}-\RRC{\hat{\omega}}=\frac{4}{\hat{\omega}\Comp}\left(\frac{1}{\hat{u}} - 1 + \frac{\Comp}{4}\hat{u}\right),
\end{equation}
where $\RRC{\hat{x}}=\dot{x}/\hat{x}$ is the relative rate of change of variable $\hat{x}(\hat{t})$, with $\dot{x}=\diff\hat{x}/\diff\hat{t}$. Steady solutions $\bar{u}$ of the reduced model are given by $\dot{u}(\bar{u})=0$. There are two steady solutions for a subcritical expulsion,
\begin{equation}\label{eq:ubar}
\bar{u}_\pm=\frac{2}{\Comp}\left(1\pm\sqrt{1-\Comp}\right),
\end{equation}
plotted as dashed and solid black lines in Fig.~\ref{fig:res3}(a). These steady solutions merge and annihilate at a saddle-node bifurcation at $\Comp=\Comp_\mathrm{crit}$, so that there are no real steady solutions for a supercritical expulsion. Physically, these steady displacements are states in which the driving pressure and the opposing resistance decrease at the same relative rate, $\RRC{\Delta\hat{p}}=\RRC{\hat{\omega}}$.

For a subcritical expulsion [Fig~\ref{fig:res3}(a)], the stability of each steady solution can be inferred from the sign of $\RRC{\hat{u}}$ close to that solution, which is indicated by the colormap in Fig.~\ref{fig:res3} (note $\hat{u}>0$). Small perturbations decay close to $\bar{u}_-$ and grow close to $\bar{u}_+$ meaning that these solutions are, respectively, an attractor and a repeller. To understand the physical mechanisms that give each branch its stability, we consider how $\RRC{\Delta\hat{p}}$ and $\RRC{\hat{\omega}}$ vary with $\hat{u}$. When $\RRC{\Delta\hat{p}}>\RRC{\hat{\omega}}$, the interface accelerates, $\RRC{\hat{u}}>0$. This is the case in most of the phase diagram: when $\hat{u}<1$ because the gas is compressing ($\Delta\dot{p}>0$ and $\dot{\omega}<0$), when $1<\hat{u}<\bar{u}_-$ because the gas is decompressing slowly ($\Delta\dot{p}<0$ and small), and when $\bar{u}_+<\hat{u}$ because the gas is overcompressed ($\Delta\hat{p}$ is large). When $\RRC{\Delta\hat{p}}<\RRC{\hat{\omega}}$, however, the interface decelerates, $\RRC{\hat{u}}<0$. This only occurs during subcritical expulsion when $\bar{u}_-<\hat{u}<\bar{u}_+$, for which the air undergoes fast decompression ($\Delta\dot{p}<0$ and large). The fast decompression region thus gives rise to the two steady states $\bar{u}_\pm$ and imparts their respective stabilities. In the supercritical phase space [Fig~\ref{fig:res3}(b)], there are only two regions because the fast decompression region collapses and ceases to exist when the two steady states annihilate at $\Comp=\Comp_\mathrm{crit}$. Hence, $\dot{u}>0$ for all $\hat{u}$.

Finally, the striking breakout features observed in our models and experiments are a consequence of the vanishing resistance at breakout. For subcritical expulsion, this means the attractive and repelling states become arbitrarily strong [$\RRC{\hat{u}}$ diverges except at $\bar{u}_\pm$] and the system terminates at $\hat{u}_{bo}\equiv\bar{u}_-$. For supercritical expulsion, in the absence of steady solutions, $\hat{u}$ diverges at the moment of breakout. To illustrate this, we plot several trajectories (magenta lines) in Figs.~\ref{fig:res3}(a, b) for a range of initial velocities $\hat{u}(0)$. In our experiments, a non-zero $\hat{u}(0)$ could be imposed by pre-compressing the air before opening a valve to initiate flow. In the subcritical phase space [Fig.~\ref{fig:res3}(a)], all trajectories with $\hat{u}(0)<\bar{u}_+$ terminate on the attractive solution $\bar{u}_-$, while those with $\hat{u}(0)>\bar{u}_+$ exhibit divergent $\hat{u}_{bo}$. In the supercritical phase space [Fig.~\ref{fig:res3}(b)], $\hat{u}$ diverges for all $\hat{u}(0)$.

In summary, we have shown that the gas-driven displacement of a confined viscous liquid is an example of a simple, first-order dynamical system (Eq.~\ref{eq:ODEred}) that exhibits unexpectedly complex dynamical regimes. A single dimensionless parameter determines whether these dynamics are subcritical or supercritical, leading to `on-time' and quasi-steady or delayed and burst-like dynamics, respectively. The key features of this idealised system are strikingly evident in our fluid-mechanical realisation, demonstrating that the underlying dynamical framework can exert a powerful control on real-world systems. We therefore expect that analogous dynamics would occur in other nonlinear systems that couple the key ingredients embodied in the reduced model: a linear spring and a reducing, state-dependent resistance. In fluid mechanical systems, for instance, spring-like compressibility can also originate from elastic walls, which act as volumetric capacitors~\citep{Guyard2022}. More broadly, analogous time-dependent currents may occur in electrical circuits, which are an established testbed of nonlinear dynamics~\citep{Healey1991}; an idealised memristor-capacitor circuit has the right key ingredients and can therefore be represented by an equation that is analogous to the reduced model. In general, subcritical dynamics could be used to mitigate the presence of spring-like components as a relatively minor and transient perturbation from corresponding ``incompressible'' behavior. Supercritical dynamics could wreak havoc if encountered unexpectedly (such as while squeezing a condiment bottle), but could also be exploited as a new design tool --- for example, as a means of detecting otherwise-imperceptible flows or currents via passive amplification.



\begin{acknowledgments}
We are grateful to Clive Baker for technical support. We thank Dominic Vella and Anne Juel for helpful discussions. This work was supported by the European Research Council (ERC) under the European Union's Horizon 2020 Programme [Grant No. 805469] and by the John Fell OUP Research Fund [Grant No. 132/012].
\end{acknowledgments}


\begin{thebibliography}{18}
\bibitem [{see()}]{see_supp}%
    \BibitemOpen
    \href@noop {} {}\bibinfo {note} {See Appendix for more information on the experimental methods, data analysis, and mathematical modelling, and which includes Refs.~[13--15].}\BibitemShut {Stop}%
\end{thebibliography}


%
  
\clearpage

\appendix

\setcounter{section}{0} \renewcommand{\thesection}{S\arabic{section}}
\setcounter{figure}{0} \renewcommand{\thefigure}{S\arabic{figure}}
\setcounter{table}{0} \renewcommand{\thetable}{S\arabic{table}}
\setcounter{equation}{0} \renewcommand{\theequation}{S\arabic{equation}}

\section{Experimental methods}

Experiments were performed in a flow cell comprising a 10~cm long glass capillary tube (Cole-Parmer) of inner radius $R=0.66\pm0.01$~mm, partially inserted horizontally through a hole in the side of an acrylic box of internal dimensions $5\times5\times10$~cm$^3$. The end of the tube external to the box was connected to a sealed reservoir of air, comprising stiff polyurethane tubing (of internal volume $3.05\pm0.05$~mL), short lengths of flexible tubing (Tygon; Saint-Gobain) and Gastight syringes (Hamilton). The total initial volume $V_i=[4, 8, 16, 32]$~mL of the air reservoir was set by adjusting the plunger of a `passive' 50~mL syringe (1050TTL; Hamilton), while an `active' 1~mL syringe (initial volume 1.00~mL; 1001TTL; Hamilton) was used to inject. The box was filled beyond the level of the capillary tube with silicone oil (Dow Corning) of dynamic viscosity $\mu=0.096$~Pa~s, surface tension $\gamma=21$~mN~m$^{-1}$, and density $\rho=960$~kg~m$^{-3}$ at laboratory temperature $T_{\mathrm{lab}}=22\pm1~^{\circ}C$. Before each experiment, oil was drawn from the box into the tube, filling an initial length $L=56\pm1$~mm; the system was allowed to equilibrate, such that the air-oil interface was stationary. The initial volume of liquid in the tubing, $\pi R^2L$, was much less than the volume of the oil bath, such that filling or draining the tube produced a negligible change in the level of the oil. Experiments were initiated by injecting air into the tube to displace oil. Injection was driven at a fixed nominal volumetric rate $Q=[0.05, 0.1, 0.2, 0.4, 0.8, 1.6]$~mL~min$^{-1}$ by depressing the plunger of the active syringe with a syringe pump (AL-4000; WPI). Variations in air temperature over the course of each experiment produced small anomalous volume fluxes $\sim 1\mu$L~min$^{-1}\ll Q$. The motion of the air-oil interface was recorded by a CMOS camera (ACA4096-30UM; Basler) at a spatial resolution of 28.9~$\mu$m~pixel$^{-1}$ and at a rate of 6.16-197 frames per second (fps), depending on $Q$. The capillary tube was back-lit using an LED array diffused through opalescent acrylic. We used image processing techniques (MATLAB 2020b) to extract from each frame the position of the advancing interface and the thickness of the oil films deposited on the walls of the tube.

To aid our analysis, we devised a method of measuring the viscous pressure drop $\Delta p$ over the length of the liquid column. The pressure at the tube outlet was assumed to be fixed at hydrostatic pressure $p_{HS}\approx90$~Pa set by the oil bath level in the box. The gauge pressure of the air reservoir $p_g$, assumed to be uniform in space, was measured with a differential pressure transducer ($\pm50$~mm~Hg; 40PC001B3A; Honeywell) attached upstream of the inlet, from which readings were taken approximately 50 times per second. Before injection, the initial gauge pressure was $p_{g0}=p_{HS}+2\gamma/R$ due to the Laplace pressure jump across the interface. During injection, the gauge pressure was $p_g=p_{HS}+\Delta p+2\gamma/b$. The radius of curvature $b$ of the interface during injection is less than the tube radius $R$, due to thin films, and was measured from experimental images. The viscous pressure drop $\Delta p=p_g-p_{g0}+2\gamma(1/R-1/b)$ can therefore be calculated indirectly from air pressure measurements, image analysis and material properties. The exact value of $p_{HS}$ is unimportant, as long as $p_{HS}\ll p_\mathrm{atm}$.

\begin{figure}
    \includegraphics[width=0.9\linewidth]{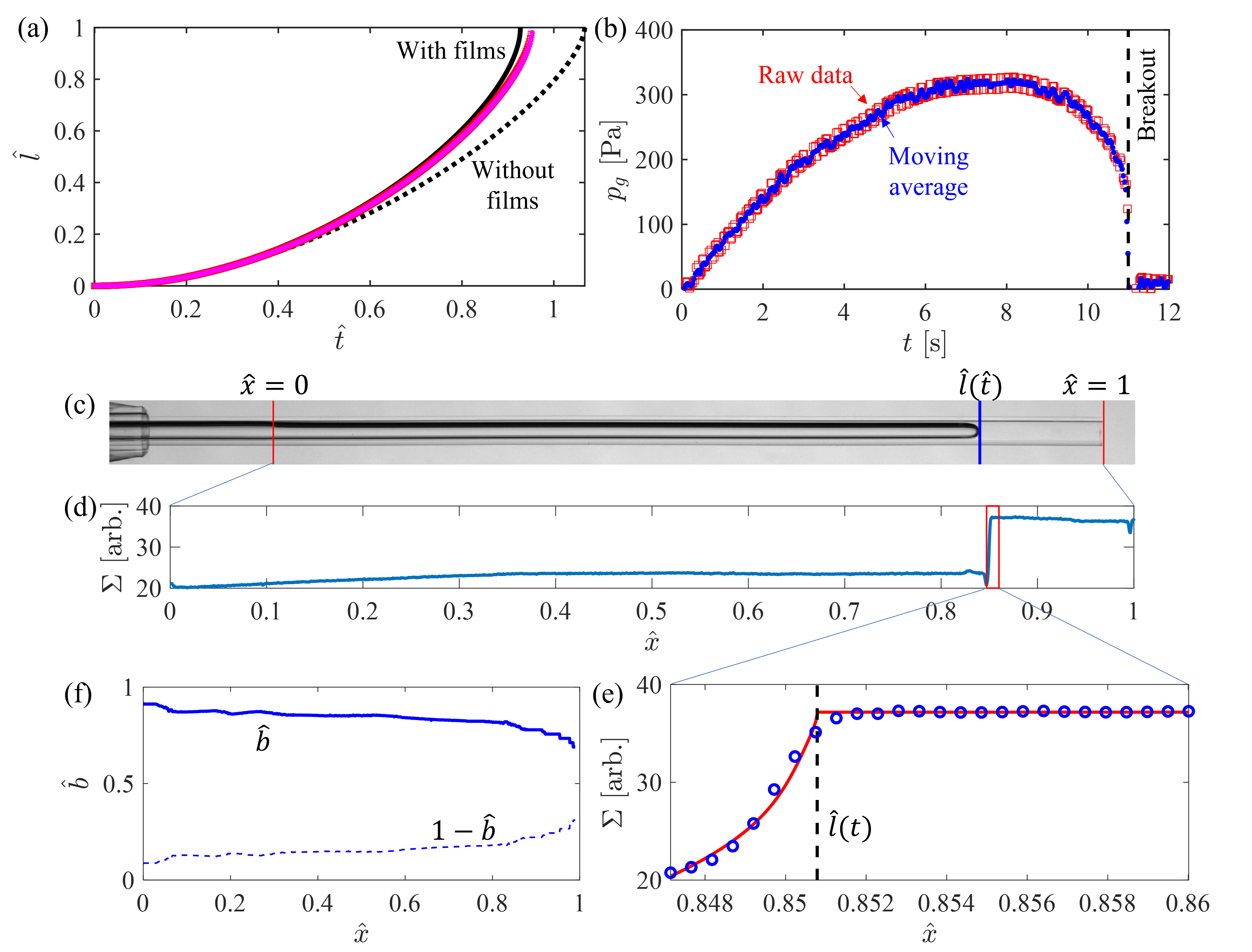}
    \caption{\label{fig:SM}Illustrations of our data analysis methods for an experiment performed at $Q=0.4$~mL/min and $V_i$=8~mL. (a)~Experimental data for normalised interface displacement $\hat{l}$ as a function of normalised time $\hat{t}$. Red squares and magenta dots show, respectively, the first and second experimental repeats at these parameters. The solid and dashed lines show, respectively, the full model with and without films (Eqs.~\ref{eq:ODEpvl} and \ref{eq:ODEnofilms}). (b)~Raw gauge pressure $p_g$ for the same experiment (red squares) as a function of time $t$. The blue dots are the processed moving-average data. The dashed line indicates the instant of breakout. (c)~A frame recorded part-way through the same experiment. The normalised axial coordinate $\hat{x}$ is indicated, as is the detected normalised interface displacement $\hat{l}$. (d-e)~Sum pixel intensity within the tube $\Sigma$ as a function of $\hat{x}$ is plotted in (d). The red box indicates the detected vicinity of the tip of the air finger, which is shown close-up in (e); here, circles are the measured values of $\Sigma$, the solid line is the fitted profile (Eq.~\ref{eq:FitFct}), and the dashed line is the detected interface displacement $\hat{l}$. (f)~Normalised radius of curvature $\hat{b}$ of the air-oil interface as a function of $\hat{x}$ (solid line). The dashed line is $1-\hat{b}$, intended to illustrate the shape of the detected air finger.}
\end{figure}

Each experiment was repeated twice. Fig.~\ref{fig:SM}(a) shows data for two experimental repeats overlaid (red squares and magenta dots), demonstrating the strong reproducibility observed. For clarity, we plot data from a single experiment for dynamical results, such as $l(t)$, $\Delta p(l)$, \textit{etc}. For breakout metrics, such as $\Delta p_{bo}$, $u_{nb}$ and $t_{bo}$, the symbols and error bars plotted correspond to the mean and standard deviation, respectively, of each pair. The main source of uncertainty in our analysis stems from the radius of the tube (known to within 10~$\mu$m), which strongly affects viscous pressures and the compressibility number.

\section{Data analysis}

\subsection{Pressure data}

Figure~\ref{fig:SM}(b) shows raw experimental data (red circles) for gauge pressure $p_g$ as a function of time $t$. To smooth out the electronic noise of the signal, we take a moving average of the data over a time-window covering one percent of the experiment's duration, from $t=0$ until breakout. In the main text, we exclusively report moving-average pressure data [blue dots in Fig.~\ref{fig:SM}(b)]. All data beyond the instant of breakout (\textit{i.e.}, $t>t(l=L)$) are discounted.

\subsection{Image analysis}

Image processing was performed in MATLAB~2020b to extract the interface displacement $l(t)$ and the radius of curvature $b$ of the air-oil interface. The main steps and results are illustrated in Figs.~\ref{fig:SM}(c-f). Fig.~\ref{fig:SM}(a) shows a typical frame from an experiment performed at $Q=0.4$~mL/min and $V_i=8$~mL. The $x$-coordinate is measured along the axis of the tube and normalised such that $\hat{x}=0$ and $\hat{x}=1$ are the initial position of the interface and the end of the tube, respectively. To locate the tip of the finger, we take advantage of the approximately circular dark region at the end of the air finger (where light is strongly refracted). We first measure $\Sigma(\hat{x})$, which is the sum of pixel intensities over each column of pixels between the tube inner walls as a function of $\hat{x}$ [Fig.~\ref{fig:SM}(d)]. We then locate the region of $\hat{x}$ where $\Sigma$ exhibits the strongest monotonic increase with increasing $\hat{x}$, assumed to be the vicinity of the tip. This region is shown in detail in Fig.~\ref{fig:SM}(e), where circles indicate the value of $\Sigma$ at each column of pixels over this range of $\hat{x}$. We fit a function of the form
\begin{equation}\label{eq:FitFct}
\Sigma = \Sigma_0 - 0.5\left[1-\tanh\left(S(\hat{x}-\hat{l})\right)\right]\sqrt{R_\mathrm{int}^2-(\hat{x}-\hat{l}+R_\mathrm{int})^2},
\end{equation}
where $\Sigma_0$, $\hat{l}$, $S$ and $R_\mathrm{int}$ are fitting parameters. Eq.~\ref{eq:FitFct} approximately matches between a region of near-uniform intensity $\Sigma_0$ for $\hat{x}>\hat{l}$ (ahead of the interface) and a region dimmed by an approximately spherical occlusion (the interface) for $\hat{x}<\hat{l}$ (behind the interface). This \textit{ad hoc} method produces reliable tracking of the interface displacement $\hat{l}(\hat{t})$. The thick blue line in Fig.~\ref{fig:SM}(c) shows the fitted value of $\hat{l}$.

To measure the radius of curvature $\hat{b}$ of the interface, we assume that the residual films do not evolve significantly over experimental timescales. In each frame at time $\hat{t}$, we subtract a background image of the experiment immediately prior to injection in order to highlight the position of the interface. We then use edge detection algorithms to locate the pixels on the interface. The radius of the air finger at each column of pixels (\textit{i.e.}, each value of $\hat{x}$) between $\hat{x}=0$ and $\hat{x}=\hat{l}(\hat{t})-2R/L$ is stored in an array; we only measure up to $\hat{x}=\hat{l}(\hat{t})-2R/L$ as we assume the films are static once the tip of the interface has advanced a sufficient distance (one tube diameter). To reduce pixel noise, the measured values of $\hat{b}$ at each position $\hat{x}$ are averaged over all recorded frames. The resulting profile of $\hat{b}(\hat{x})$ for the experiment performed at $Q=0.4$~mL/min and $V_i$=8~mL is shown in Fig.~\ref{fig:SM}(f). There is more pixel noise closer to the outlet at $\hat{x}=1$ because these films form later, meaning there are less recorded frames over which to average the measured values of $\hat{b}$. For reference, we also plot $1-\hat{b}$ (dashed line) to give a sense of the shape of the air finger. Nonmonotonic variations in $\hat{b}$ suggest the uncertainty of this method is around $\pm$~10~$\mu$m (around 1.5\% of $R$). We expect refractive distortions to be weak due to the similar refractive indices of the glass tube and the silicone oil, as well as due to the small thickness of the tube walls (0.3~mm).

\subsection{Interface velocity}

The interface velocity $\hat{u}=\diff \hat{l}/\diff \hat{t}$ is measured from experimental data by taking least-squares linear fits to segments of $\hat{l}(\hat{t})$ data. The value of $\hat{u}$ at time $\hat{t}$ is taken to be the gradient $m$ of the linear fit $\hat{l}_i=m\hat{t}_i+c$ for data $\hat{l}_i$ and $\hat{t}_i$ in the range $(\hat{t}-\Delta \hat{t})<\hat{t}_i<(\hat{t}+\Delta \hat{t})$, where $\Delta \hat{t}$ is 1.43\% of the total normalised duration of the experiment. At lower values of $Q$, there is evidence of stick-slip behaviour causing non-smooth motion of the syringe pump, though the effect was not severe enough to alter or cloud the conclusions of our study.

\section{Mathematical modelling}
\subsection{Full model with thin films}
The mathematical model describing compressible displacement in the absence of thin residual films is given by Eq.~4 of the main text,
\begin{equation}\label{eq:ODEnofilms}
\frac{\diff \hat{l}}{\diff \hat{t}}=\left[\frac{\hat{p}_0+2/\hat{\mathrm{Ca}}}{\hat{V}_i+\left(\hat{l}-\hat{t}\right)}\right]\left(\frac{\hat{t}-\hat{l}}{1-\hat{l}}\right),
\end{equation}
where we use dimensionless variables for clarity. In experiments, the tube walls are coated in a residual liquid film of thickness $\hat{\delta}_{\mathrm{film}}=\delta_{\mathrm{film}}/R$. We assume that $\hat{\delta}_{\mathrm{film}}$ depends on the local capillary number $\hat{\mathrm{Ca}}_l=\mu Q/(\pi R^2 \gamma)$ and the dimensionless velocity $\hat{u}=\diff\hat{l}/\diff\hat{t}$, via
\begin{equation}\label{eq:tFilm}
\hat{\delta}_\mathrm{film}=\frac{A\left(\Cah_l\hat{u}\right)^{2/3}}{1+B\left(\Cah_l\hat{u}\right)^{2/3}}.
\end{equation}
The form of Eq.~\ref{eq:tFilm}, proposed by~\citet{Aussillous2000}, reduces to the classical~\citet{Bretherton1961} law in the limit of low $\hat{\mathrm{Ca}}_l$, while saturating at a finite thickness at high $\hat{\mathrm{Ca}}_l$ due to the confining geometry of the tube. We choose to use the values $A=1.34$ and $B=3.74$ derived by~\citet{Klaseboer2014} which produce excellent agreement with our experimental data. Thin films effectively lead the air to propagate in a narrower tube of radius
\begin{equation}\label{eq:BubRad}
\hat{b}=1-\hat{\delta}_{\mathrm{film}}.
\end{equation}
This has two key effects on the dynamics: (i) The tip of the air finger is squeezed, leading to a greater capillary pressure drop $p_c=2\gamma/b$ across the interface; (ii) The kinematic boundary condition is modified by volume conservation such that the interface velocity $\hat{u}=\diff\hat{l}/\diff\hat{t}=\hat{U}/\hat{b}^2$ is greater than the mean velocity $\hat{U}=\Delta \hat{p}/(1-\hat{l})$ of the liquid slug downstream. The full model with thin films is then
\begin{subequations}\label{eq:ODEfilms}
\begin{equation}
\frac{\diff \hat{l}}{\diff \hat{t}}=\left[\frac{2\hat{V}_i/\hat{\mathrm{Ca}}\left(1-\frac{1}{\hat{b}}\right)+\left(\hat{p}_0+\frac{2}{\hat{\mathrm{Ca}}\hat{b}}\right)\left(\hat{t}-\hat{V}_b\right)}{\hat{V}_i + \hat{V}_b - t}\right]\frac{1}{\hat{b}^2\left(1-\hat{l}\right)}=\frac{\Delta\hat{p}}{\hat{b}^2\left(1-\hat{l}\right)},~\mathrm{and}
\end{equation}
\begin{equation}\label{eq:dVbdt}
\frac{\diff \hat{V}_b}{\diff \hat{t}}=\hat{b}^2\frac{\diff \hat{l}}{\diff \hat{t}}.
\end{equation}
\end{subequations}
where $\hat{V}_b$ is the dimensionless volume of liquid expelled at time $\hat{t}$.

To avoid the divergent breakout velocities present in the high-compressibility regime, we recast the full model in terms of the pressure drop $\Delta \hat{p}$ versus interface position $\hat{l}$, which gives
\begin{subequations}\label{eq:ODEpvl}
\begin{equation}
\frac{\diff \left(\Delta \hat{p}\right)}{\diff \hat{l}}=\frac{2}{\hat{b}^2\Cah}\frac{\diff \hat{b}}{\diff \hat{l}}+\left[\frac{\hat{V}_i\left(\hat{p}_0+\frac{2}{\Cah}\right)}{\left(\hat{V}_i+\hat{V}_b-\hat{t}\right)^2}\right]\left(\frac{\hat{b}^2\left(1-\hat{l}\right)}{\Delta\hat{p}}-\frac{\diff \hat{V}_b}{\diff \hat{l}}\right),
\end{equation}
\begin{equation}
\frac{\diff \hat{b}}{\diff \hat{l}}=\frac{\frac{2}{3}A\left(\Cah_l\hat{u}\right)^{2/3}}{\hat{u}\left(1+B\left(\Cah_l\hat{u}\right)^{2/3}\right)}\frac{\diff \hat{u}}{\diff \hat{l}}\left(\frac{B\left(\Cah_l\hat{u}\right)^{2/3}}{1+B\left(\Cah_l\hat{u}\right)^{2/3}}-1\right),~\mathrm{and}
\end{equation}
\begin{equation}
\frac{\diff \hat{V}_b}{\diff \hat{l}}=\hat{b}^2,
\end{equation}
\end{subequations}
where
\begin{subequations}\label{eq:tNv}
\begin{equation}
\hat{t}=\hat{V}_b+\hat{V}_i\left(\frac{\Delta\hat{p}+\frac{2}{\hat{b}\Cah}-\frac{2}{\Cah}}{\Delta\hat{p}+\frac{2}{\hat{b}\Cah}+\hat{p}_0}\right),~\mathrm{and}
\end{equation}
\begin{equation}
\frac{\diff \hat{u}}{\diff \hat{l}}=\frac{1}{\hat{b}^2\left(1-\hat{l}\right)}\left[\frac{\diff \left(\Delta\hat{p}\right)}{\diff \hat{l}} - \left(\frac{2\Delta\hat{p}}{\hat{b}}\right)\frac{\diff \hat{b}}{\diff \hat{l}} + \frac{\Delta\hat{p}}{\left(1-\hat{l}\right)}\right].
\end{equation}
\end{subequations}
Equation~\ref{eq:ODEpvl} is a system of nonlinear implicit ordinary differential equations, which we solve with the built-in function ODE15I in MATLAB 2020b, subject to the initial conditions $\hat{l}(0)=0$, $\Delta\hat{p}(0)=10^{-10}$, $\hat{b}(0)=1$, $\hat{V}_b(0)=0$, and $(d\hat{V}_b/d\hat{l})(0)=1$. To reduce the number of steps needed for the solver to converge onto the solution, we first solve Eq.~\ref{eq:ODEfilms} with $\hat{u}(0)=0$, then take the values of $d(\Delta\hat{p})/d\hat{l}$ and $d\hat{b}/d\hat{l}$ at the first nonzero timestep to be the initial values for Eq.~\ref{eq:ODEpvl}. In Fig.~\ref{fig:SM}(a), we plot $\hat{l}(\hat{t})$ from the full model both with films (solid line; Eq.~\ref{eq:ODEpvl}) and without films (dashed line; Eq.~\ref{eq:ODEnofilms}) for comparison. Both produce qualitatively similar dynamics, but the model with films gives much better quantitative agreement with experiments.

\subsection{Reduced model: Analytical implicit solution}
The reduced model [Eq.~\eqref{eq:ODEred} in the main text] permits the implicit analytical solution
\begin{equation}\label{eq:ImpSol1}
\ln{\left[\left(\hat{l}-1\right)^2+\frac{4\left(\hat{t}-\hat{l}\right)\left(\hat{t}-1\right)}{\Comp}\right]}+\frac{2}{\sqrt{\Comp-1}}\left[\arctan{\left(\frac{1}{\sqrt{\Comp-1}}\right)} - \arctan{\left(\frac{2\hat{t}-\hat{l}-1}{\left(\hat{l}-1\right)\sqrt{\Comp-1}}\right)}\right]=0
\end{equation}
when $\Comp>1$, or
\begin{equation}\label{eq:ImpSol2}
\ln{\left[\left(\hat{l}-1\right)^2+\frac{4\left(\hat{t}-\hat{l}\right)\left(\hat{t}-1\right)}{\Comp}\right]}-\frac{2}{\sqrt{1-\Comp}}\left[\arctanh{\left(\frac{1}{\sqrt{1-\Comp}}\right)} - \arctanh{\left(\frac{2\hat{t}-\hat{l}-1}{\left(\hat{l}-1\right)\sqrt{1-\Comp}}\right)}\right]=0
\end{equation}
when $\Comp\le1$. The breakout time $\hat{t}_{bo}>1$ in the high-compressibility regime ($\Comp>1$) can be found by substituting $\hat{l}=1$ into Eq.~\ref{eq:ImpSol1}, which implies that the breakout velocity $\hat{u}_{bo}$ is divergent in this regime (note that $\hat{u}_{bo}$ is measured \textit{at} breakout, as opposed to the near-breakout velocity $\hat{u}_{nb}$ discussed in the main text). Similarly, the breakout velocity in the low-compressibility regime ($\Comp\le1$) can by found by taking $\hat{l}=1-\epsilon$ and $\hat{t}_{bo}=1+\epsilon(\Comp \hat{u}_{bo} - 1)$ (from Eq.~\ref{eq:ODEred}) in the limit $\epsilon\to0$ in Eq.~\ref{eq:ImpSol2}, which gives $1\le\hat{u}_{bo}\le2$. The finite breakout velocities in the low compressibility regime imply that the breakout time $\hat{t}_{bo}=1$. Finally, the breakout pressure drop is given by $\Delta\hat{p}_{bo}=4(\hat{t}_{bo}-1)/\Comp$, from which it is clear that a delayed breakout ($\hat{t}_{bo}>1$) must be accompanied by an over-pressure ($\Delta\hat{p}_{bo}>0$). Table~\ref{tab:BO} summarises these results.

\begin{table}[hb]
    \caption{Analytical results for the times, velocities, and pressures and breakout in the low- and high-compressibility regimes ($\Comp\le1$ and $\Comp>1$, respectively).}
    \centering
    \begin{tabular}{r c c c}
        & $\Comp\le1$ &$~~~~~$& $\Comp>1$ \\ [0.5ex]
        \hline
        $\hat{t}_{bo}$ \vline& 1 & & $1+\sqrt{\Comp/4}\exp\left[\frac{-1}{\sqrt{\Comp-1}}\left(\frac{\pi}{2}+\arctan\left(\frac{1}{\sqrt{\Comp-1}}\right)\right)\right]$ \\
        $\hat{u}_{bo}$ \vline & $1\le\frac{2}{\Comp}\left(1 - \sqrt{1-\Comp}\right)\le2$ & & $\infty$ \\
        $\hat{p}_{bo}$ \vline & 0 & & $\frac{1}{\sqrt{\Comp/4}}\exp\left[\frac{-1}{\sqrt{\Comp-1}}\left(\frac{\pi}{2}+\arctan\left(\frac{1}{\sqrt{\Comp-1}}\right)\right)\right]$ \\ [1ex]
        \hline
    \end{tabular}\label{tab:BO}
\end{table}

\end{document}